\documentclass[twoside]{article}
\usepackage{graphicx}

\begin{document}

\title{Quasi-phase-matched high harmonic generation in corrugated micrometer-scale waveguides}

\author{A. Husakou\\
{Max Born Institue, Max Born Str. 2a, D-12489 Berlin} \\
{gusakov@mbi-berlin.de} }

\maketitle

\begin{abstract}
The high harmonic generation in periodically corrugated submicrometer waveguides is studied numerically. Plasmonic field enhancement in the vicinity of the corrugations allows to use low pump intensities. Simultaneously, periodic placement of the corrugations leads to quasi-phase-matching and corresponding increase of the high harmonic efficiency. The optimization of waveguide geometry is performed, and the resulting spectra are analyzed by the means of (1+1)D numerical model.
\end{abstract}

\section{Introduction}

High-harmonic generation (HHG) is the key light source  in the extreme ultraviolet region. One of the most important applications of the HHG is the generation of attosecond pulses\cite{hent}, which permits subfemtosecond-resolution studies of fundamental electronic processes in atoms and molecules and provides a valuable tool for attosecond molecular physics, nanolithography, et al. \cite{corkum,paul,krausz,sansone}. The main mechanism of HHG is described by the three-step model, in which an electron undergoes ionization, acceleration in the strong electric field, and recombination accompanied by the emission of high-energy photons \cite{Lewenstein}. 

Recently the generation of high harmonics in gases in the vicinity of metal nanostructures (nanostructrure-assited high-harmonic generation, NA-HHG) \cite{kim} and micrometer-sized hollow metallic fibers\cite{kim2} was experimentally studied. Such setup allows to use significantly reduced pump intensities due to plasmonic field enhancement in the vicinity of nanostructures. A theoretical investigation of HHG using field enhancement by metallic nanostructures was provided by Ref. \cite{Husakou} where the Lewenstein model was extended to incorporate the field inhomogeneity in the hot spots and electron absorption from the metal surfaces. After that, in a number of publications the quantum description of both the field inhomogeneity and electron absorption at surface was provided by several groups \cite{Ciappina1,Ciappina2,Yavuz,ciap1,shaa1,yav,fet,cia2,shaa2,cia3}, along with the studies of the high-energy photon emission in plasmonic field \cite{perez}, plasmon-assisted as-pulse generation without CEO-stabilization \cite{luo}, HHG in periodic bowtie arrays \cite{pfull}, and theoretical studies of NA-HHG employing nanoplasmonic field enhancement of coupled ellipsoids \cite{Stebbings}. Besides, NA-HHG in the vicinity of a rough metallic surface \cite{rough}, in an array of coupled metal nanospheres \cite{spheres}, and the generation of isolated attosecond pulses using polarization gating by crossed bowtie nanostructures \cite{Husakou2} have been investigated. Quite notably, experimental investigation of NA-HHG in tapered hollow metallic waveguides \cite{kim2,njp} demonstrated the possibility to achieve high harmonic numbers using low-intensity pump.

A typical disadvantage of NA-HHG is the low efficiency ($\sim$10$^{-9}$) of the spectral transformation due to the short propagation length, compared to typical efficiencies in noble gases of 10$^{-6}$. In contrast to other systems, the NA-HHG efficiency is difficult to increase by increasing the propagation length due to significant phase mismatch. Besides, the approach of Ref. \cite{kim} was criticized in Ref. \cite{ropers}, which interprets the experimental peaks as fluorescence peaks from the noble-gas atoms rather than as high harmonics. This controversy apply only to set-ups with rather low HHG efficiencies like in \cite{kim}. Therefore it is desirable to investigate the possibilities to increase the efficiency of the NA-HHG. 

The key factor which limits the  nanostructure-assisted HHG efficiency is a large phase mismatch.  It can be overcome by periodic modulation of the medium properties, so-called quasi-phase-matching. Previously, in the context of standard HHG in gases, several approaches \cite{counter,nozzles,ganeev} were proposed and implemented for this aim. In the context of NA-HHG, quasi-phase-matching in a metal nanoparticle-gas composite \cite{ourqpm} was proposed. However, in view of the development of the HHG generation in very thin waveguides \cite{kim2,njp}, it is important to investigate the setups to increase the HHG efficiency in the framework of the waveguide design.

 In the context of the NA-HHG, the most suitable of these approaches is the corrugation of the waveguide: the periodic change of the waveguide radius leads to the modulations of the pump beam intensity with a predesigned period. Since the HHG strongly nonlinearly depends on the pump intensity, this results in the quasi-phase-matching. This approach was used in several experimental and theoretical publications for the standard HHG \cite{corrug,corrug2}.  Here we investigate quasi-phase-matched NA-HHG waveguides with radii comparable or below the wavelength. The corrugation (localized reduction of the inner radius) of such waveguides will result in pump field enhancement, leading to lowered pump-intensity requirements. By placing the corrugations at the specific period, one can expect to achieve the quasi-phase-matching between the pump wave and a high harmonic of a certain order.

In this paper, we investigate such a design using the first-principle analysis of the modal field and its enhancement, accompanied by the (1+1)D model for the pump and harmonic field propagation. We show that one can indeed achieve the quasi-phase matching in such a system, leading to the increased efficiency for a chosen harmonics. The optimization of the waveguide geometry is performed.

The paper is organized as follows. In Section 2 we present the general setup, the distribution of the modal field and its enhancement as the function of the waveguide geometrical parameters. In Section 3, the numerical approach is presented in detail. In Section 4, we present and discuss the output harmonics spectra as well as the predicted dependence of the efficiency on the propagation length for quasi-phase-matched and non-quasi-phase-matched situations.

\section{Proposed setup}

\begin{figure}[htbp]
\centering
\fbox{\includegraphics[width=0.9\linewidth]{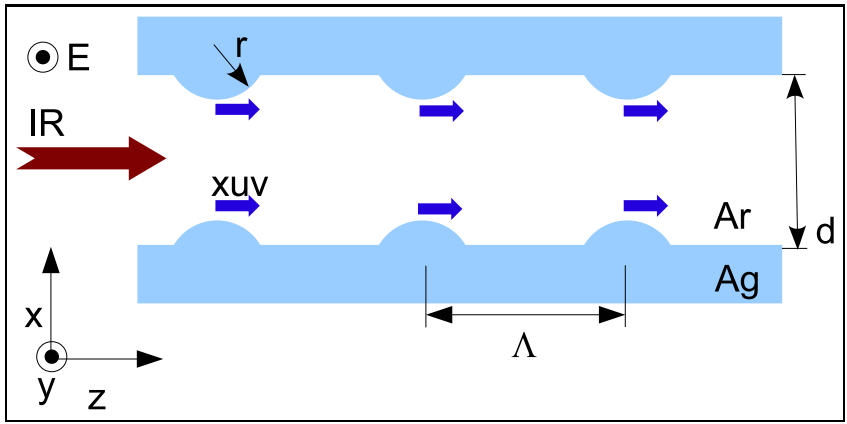}}
\caption{The scheme of the proposed setup. The $y$-polarized near-IR beam passes through an Argon-filled, silver waveguide. The waveguide geometry is characterized by the radius of the corrugations $r$, the distance between the waveguide walls $d$, and the corrugation periodicity $\Lambda$. The structure is considered infinite in $y$ direction, the thickness of the waveguide walls is 50 nm, the curvature center of the corrugations is in the middle of the wall, and xuv high harmonics are emitted in the vicinity of the corrugations as depicted.}
\label{scheme}
\end{figure}

In Fig. \ref{scheme}, a scheme of the proposed setup is presented. We consider a plane waveguide (invariant in $y$ direction) which consists of two metal slabs with thickness of 50 nm each, separated by the distance $d$. Such plane geometry of the waveguide can be more easily manufactured experimentally (e.g. by lithography) than a waveguide with a circular hole. The thickness of 50 nm is larger than the skin layer depth and is sufficient to localize the pump light in the inner core of the waveguide. The corrugations consist of arc-shaped metal areas with radius $r$ separated by distance $\Lambda$, which protrude inside the waveguide, as depicted in Fig. \ref{scheme}. The $y$-polarized near-IR pump enters the waveguide from the left. Excitation of localized plasmons and associated field enhancement occur in the vicinity of the corrugations. The high harmonics which are emitted in gas propagate to the right through the waveguide core, adding up constructively in the case of the quasi-phase-matching.

\begin{figure}[htbp]
\centering
\fbox{\includegraphics[width=0.9\linewidth]{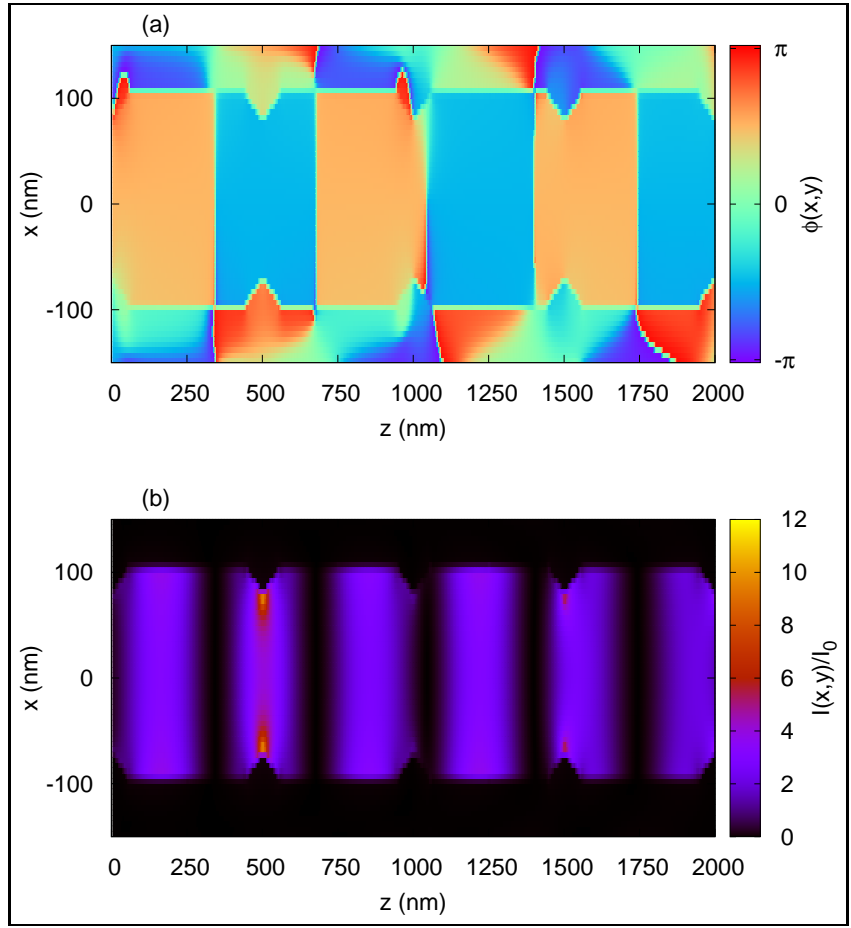}}
\caption{The distribution of the phase of the pump field (a) and of the square of the pump field (b) inside the waveguide. The parameters of the waveguide are $r$ = 75 nm, $d$ = 250 nm, the pump wavelength is 800 nm, the distance between the corrugations $\Lambda$ is 500 nm.}
\label{maps}
\end{figure}

Before presenting the nonlinear numerical model for the propagation of the pump beam and the harmonics, let us analyze the linear properties of such a  waveguide. 
The linear propagation of pump light was analyzed using the JCMwave commercial software, which provides rigorous solution of the Maxwell equations, with silver properties being properly described by both real and imaginary parts of the dielectric function. In Fig. \ref{maps}(b) one can see the typical distribution of the light intensity inside the waveguide for the cw pump wave. Note that the periodic oscillations of the field amplitude in the $z$ direction are due to the resolved optical wavelength. In the vicinity of the corrugations, the intensity is enhanced by a factor up to 15. Also, the intensity of the pump reduces with the propagation due to the modal loss of the waveguide. In Fig. \ref{maps}(a), the spatial map of the phase of the field is presented. One can see the oscillations of the phase with period equal to the wavelength in the waveguide, which is different from the free-space wavelength due to the waveguide contribution to dispersion, as well as change of the phase at the metal interface. From the data presented in Fig. \ref{maps} both real and imaginary parts of the effective refractive index of the waveguide were extracted. Note that here and hereafter we neglect the energy transfer between the transverse modes, since considered propagation distances are small compared to the distance over which such transfer takes place.

\begin{figure}[htbp]
\centering
\fbox{\includegraphics[width=0.8\linewidth]{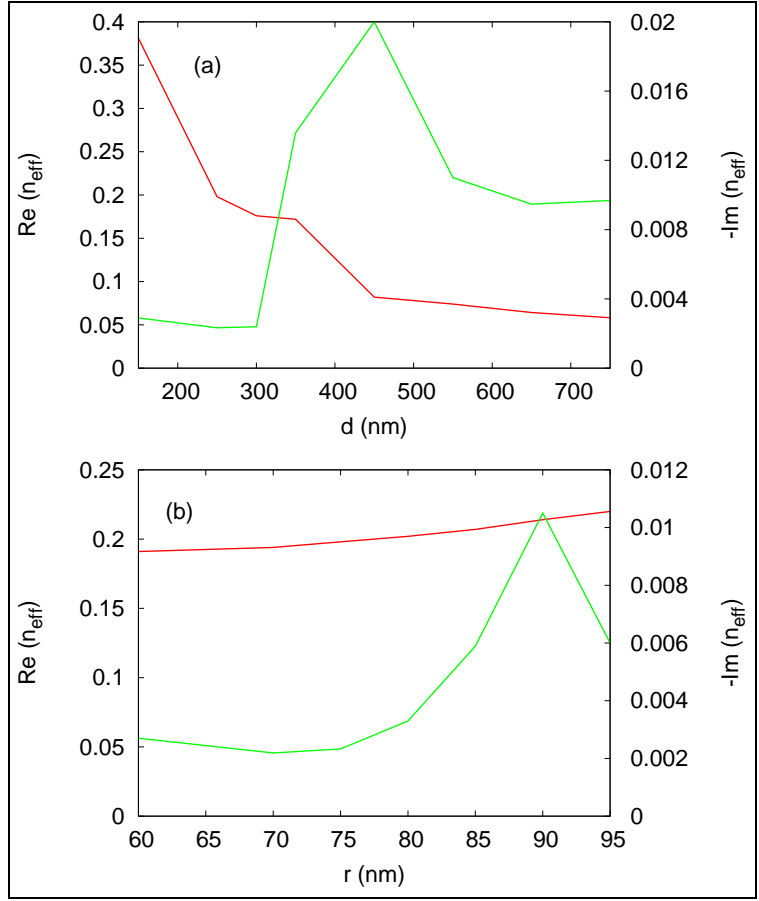}}
\caption{The dependence of the waveguide contribution to effective refractive index of the waveguide on the core size $d$ (a) and the corrugation radius $r$ (b). The real part is shown by the red curves (left axis),  the sign-reversed imaginary part is shown by the green curves (right axis), the wavelength is 800 nm, $\Lambda=300$ nm. In the case of (a), $r=75$ nm; in the case of (b), $d=250$ nm.}
\label{deps}
\end{figure}

In Fig. \ref{deps} the dependence of the effective refractive index of the waveguide on its geometrical parameters is presented. One can see that both reduction of the core size and increase of the corrugation lead to the increase of the real part of the $n_{eff}$, which is caused by the tighter light localization in the $x$ direction. The numerical dependence of the imaginary part of the refractive index, which corresponds to the modal losses, is more complicated. It exhibits minima at $d=250$ nm and 650 nm, and is almost independent on the corrugation size for $r$ up to 85 nm. Based on the presented data, we have chosen the case $d=250$ nm and $r=75$ nm for further study. It is characterized by relatively low loss as well as relatively low real part of $n_{eff}-1$, which means that harmonics with higher order can be quasi-phase-matched.

\section{Numerical model}

Let us now present the numerical model which was used to simulate the propagation of pump pulse propagation and the harmonic pulse. 

The numerical treatment of the problem under consideration includes three components: microscopic description of the HHG process, propagation of the pump pulse, and the propagation of the generated harmonics. The harmonic generation was modeled using the strong-field formalism developed by Lewenstein {\it et al.} \cite{Lewenstein}, using the following expression for the dipole moment:
\begin{eqnarray}
d_{HH}(z,t)&=&S(z)\frac{ie}{2\omega_0^{5/2}m_e}\int_{-\infty}^t\left(\frac{\pi}{\epsilon+i(t-t_s)}\right)^{3/2}\times\cr &&d(p_{ps}-eA(z,t_s))d^*(p_{ps}-eA(z,t))E(t_s)\times\cr &&\kappa(z)e^{\frac{iS(t,t_s)}{\hbar}}dt_s+c.c.
\label{dhh}
\end{eqnarray}
where $\omega_0$ is the pump frequency, $d(p)$ is the dipole moment of the outer-shell orbital, $p_{st}$ is the saddle-point canonical momentum, $A(z,t)$ is the vector-potential of the enhanced electric field $\kappa(z)E(t)$. The enhancement $\kappa(z)$ is  equal to the maximum along $x$ direction of the plasmonic enhancement inside the waveguide at the position $z$. However, one should also take into account that only a part of the transverse cross-section is characterized by this high enhancement. This is accounted for by mulitplying the harmonic dipole moment by the pre-factor $S(z)$, which is $x$-direction full-width half-maximum  of the enhancement divided by the waveguide width $d$. The classical action is defined by  $S(t,t_s)=\int_{t_s}^t(I_p-\frac12[p-eA(t')]^2/m_e)dt'$ with $I_p$ being the ionization potential. No approximation of zero velocity of the ionized electron was used, which allowed for the accurate description of the spectra also well below the harmonic threshold. Analytic expressions for dipole transition moments of the hydrogen-like atoms \cite{Lewenstein}  were utilized. 

For the propagation of the pump pulse we have numerically solved a (1+1)D unidirectional propagation equation:

\begin{equation}
\frac{\partial E}{\partial z}=i\beta(\omega)E(z,\omega)+\frac{i\omega^2}{2c^2\epsilon_0\beta(\omega)}P_{NL}(z,\omega)
\label{main}
\end{equation}

where $\beta(\omega)=n_{eff}(\omega)\omega/c$ is the wavenumber and $P_{NL}$ is the nonlinear polarization, determined by

\begin{equation}
P_{NL}=\epsilon_0\chi_3E(z,t)
\label{pnl}
\end{equation}

where $\chi_3$ is the third-order susceptibility. We note that the third-order susceptibility has a very minor effect on the propagation. The plasma contribution to the refractive index was neglected, since the corresponding change of the refractive index is significantly lower than the waveguide contribution to the effective refractive index.
 Finally, the propagation of the high-order harmonics is governed by the (1+1)D propagation equation, which incorporates a high-harmonic polarization using harmonic source dipole moment $d_{HH}(z,t)$ calculated as described above, as well as harmonic reabsorption by the argon gas. For further details of the numerical treatment see e.g. \cite{ganeev}.

\section{Results and discussion}

\begin{figure}[htbp]
\centering
\fbox{\includegraphics[width=0.9\linewidth]{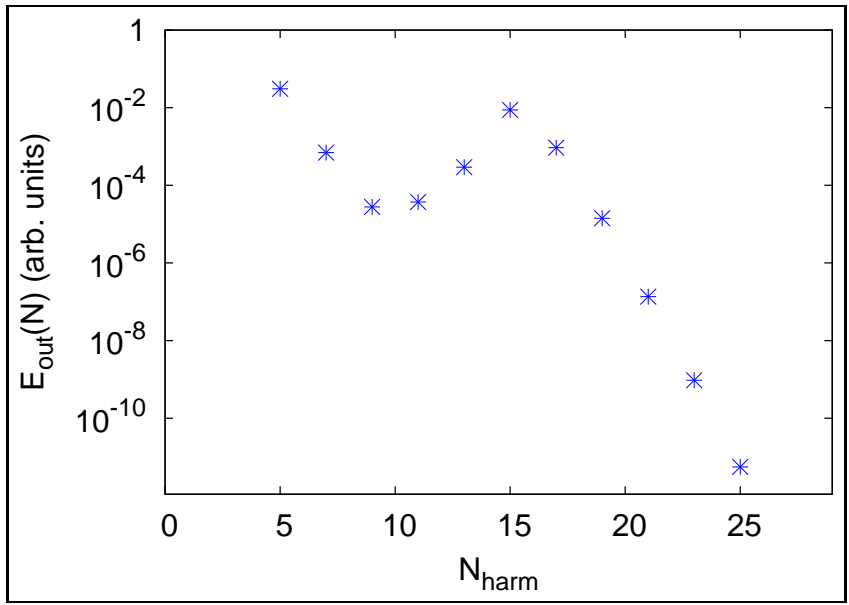}}
\caption{The output spectrum for the 30-fs FWHM pump pulse with intensity of 3.5 TW/cm$^2$ after propagation of 5 $\mu$m in a waveguide with parameters $d=250$ nm, $r=75$ nm, $\Lambda=0.3$ $\mu$m, filled with argon.}
\label{spectrum}
\end{figure}

In Fig. \ref{spectrum}, the output harmonic spectrum is shown for the input pulse and waveguide parameters listed in the caption. The period of the modulation of $\Lambda=0.3$ $\mu m$ corresponds to the first-order quasi-phase matching length of the 15th harmonic, given by $\lambda_0/[15(Re(n_{eff}-1)]$. Therefore one can observe a pronounced maximum in the spectrum for the 15th harmonic, with contrast of roughly two orders of magnitude. Note that the predicted cutoff at 23rd harmonic corresponds to the enhanced intensity of 40 TW/cm$^2$. This intensity is in agreement with the pump intensity of 3.5 TW/cm$^2$ and the maximum enhancement factor of around 10.

\begin{figure}[htbp]
\centering
\fbox{\includegraphics[width=0.9\linewidth]{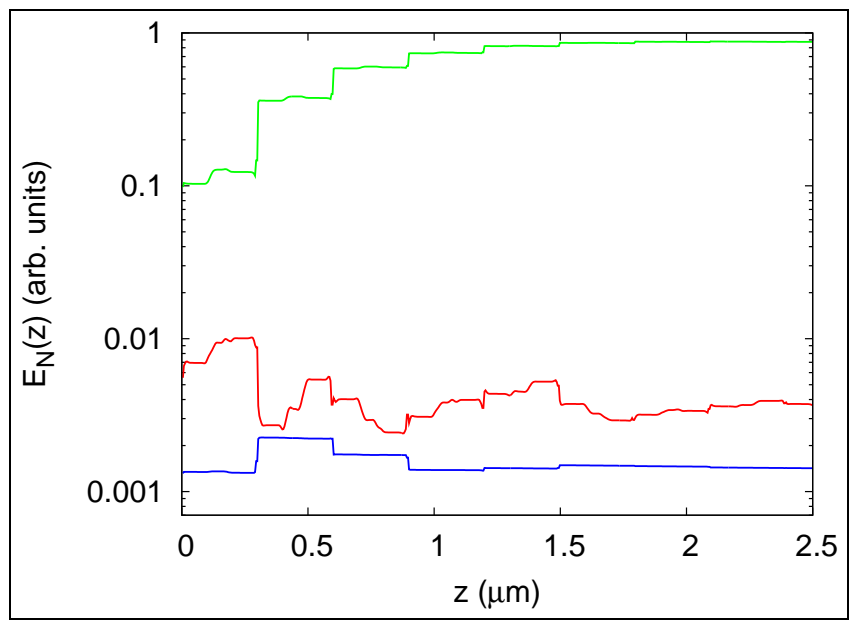}}
\caption{The evolution of the harmonic energy with he propagation length for 11th (green), 15th (blue), and 19th (red) harmonic. Other parameters are the same as in Fig. \ref{spectrum}.}
\label{evol}
\end{figure}

To confirm the quasi-phase-matching in the considered case, in Fig. \ref{evol} we present the evolution of the harmonic energy with propagation for different harmonic numbers. One can see that for the quasi-phase-matched 15th harmonic, all the contributions from the different corrugations add up constructively. This leads to a steady growth of the energy proportional to the square of the propagation distance, which overgoes into a decay after the waveguide loss reduces the pump intensity and harmonic reabsorption comes into play. On the other hand, the evolution of the energy of the 11th and the 19th harmonics shows incoherent addition of contributions from different corrugations, since the corrugation period does not correspond to the quasi-phase-matching period for these harmonics.

\section{Conclusion}
We have predicted that quasi-phase-matching in nanometer-scale corrugated hollow waveguides can be used to enhance the efficiency of high harmonic generation. The rigorous analysis of the modal properties of such waveguides was utilized together with the (1+1)D model for the pump beam and harmonics propagation. The optimization of the corrugated waveguide geometry is performed.

\section{Funding Information}

German Research Council (HU 1593/2-1).

\end{document}